\documentclass{article}
\usepackage[english]{babel}

\usepackage[letterpaper,top=2cm,bottom=2cm,left=3cm,right=3cm,marginparwidth=1.75cm]{geometry}

\usepackage{subcaption}
\usepackage{amsmath,amssymb,amsfonts,authblk}
\usepackage{graphicx}
\usepackage{tikz}
\usetikzlibrary{shapes,arrows}
\usetikzlibrary{quantikz2}
\usepackage{amsmath} 
\usepackage{listings}
\usepackage{xcolor}
\usepackage{tikz}
\usepackage{algorithm}
\usepackage{booktabs}
\usepackage{algorithmicx}
\usepackage{algpseudocode}
\usetikzlibrary{arrows.meta, positioning}
\usepackage[normalem]{ulem}
\usepackage{pgfplots}
\pgfplotsset{compat=1.18} % or 1.17 depending on your TeX distribution
\usepackage[colorlinks=true, allcolors=blue]{hyperref}

\definecolor{myBlue}{RGB}{0,133,255}

\usepackage[colorinlistoftodos,prependcaption,textsize=tiny]{todonotes}

\title{Quantum Computing for Healthcare Digital Twin Systems} 
\author[1]{Asma Taheri Monfared}
\author[1]{Andrea Bombarda}
\author[1]{Angelo Gargantini}
\author[2]{Majid Haghparast}

\affil[1]{University of Bergamo, Bergamo, Italy}
\affil[2]{Faculty of Information Technology, University of Jyväskylä, Finland}

\date{}

\begin{document}

\maketitle

\begin{abstract}
The growing complexity of healthcare systems requires advanced computational models for real-time monitoring, secure data exchange, and intelligent decision-making. Digital Twins (DTs) provide virtual representations of physical healthcare entities, enabling continuous patient monitoring and personalized care. However, classical DT frameworks face limitations in scalability, computational efficiency, and security.
Recent studies have introduced Quantum Digital Twins (QDTs) to enhance performance through quantum computing, addressing challenges such as quantum-resistant security and efficient task offloading in healthcare environments. Despite these advances, most existing QDT models remain constrained by fundamental challenges related to quantum hardware limitations, hybrid classical–quantum system integration, cloud-based quantum access, scalability, and clinical trust. 
This paper provides a comprehensive review of QDTs for healthcare, with a particular focus on identifying and analyzing the key challenges that currently hinder their real-world adoption. Furthermore, it outlines critical research directions and enabling strategies aimed at advancing the development of secure, reliable, and clinically viable quantum digital twin systems for next-generation healthcare applications.
\end{abstract}


\section{Introduction}
Digital Twin (DT) technology has emerged as a key paradigm for creating virtual representations of physical systems that are continuously updated through data exchange, enabling monitoring, simulation, and data-driven decision-making~\cite{bersani2022engineering}. 
In recent years, the healthcare sector has increasingly explored the adoption of DT technologies to support medical care, patient treatment, and device development. In healthcare, DTs are increasingly explored to improve medical care and patient treatment, enabling real-time decision support, risk assessment, and system evaluation through continuously updated models that represent patients, medical devices, and clinical processes~\cite{ahmed2021potential, jimenez2019health}. DTs have also been investigated as platforms for training medical personnel and improving preparation for clinical procedures, with reported benefits including early detection of medical conditions, evaluation of treatment alternatives, and enhanced surgical planning~\cite{ahmed2021potential, jimenez2019health}. These opportunities are aligned with broader evidence that DT-driven healthcare applications can integrate real-time data, analytics, and modeling to support patient care, predictive decision-making, workflow optimization, and training and simulation~\cite{vallee2023digital}.
Despite these promising opportunities, several challenges continue to limit the widespread adoption of DTs in healthcare. Beyond technological and methodological concerns, major obstacles arise from the inherent complexity of modeling human physiology, behavior, and clinical workflows, as well as from the presence of uncertainty in medical data and processes. Furthermore, the handling of sensitive patient information raises critical issues related to privacy and data protection. More broadly, the effective exploitation of DTs depends on the level of trust that stakeholders place in the models and the insights they provide. Existing studies have focused mainly on security and privacy aspects~\cite{fuller2020digital}, while challenges related to computational scalability, latency, and real-time analytics remain insufficiently addressed in healthcare DT systems.

These limitations motivate the investigation of advanced computing paradigms capable of supporting the increasing complexity and performance requirements of healthcare DTs. In this context, quantum computing has recently been explored as a promising enabler for next-generation digital twin architectures. Recent studies indicate that quantum-assisted approaches can help overcome key limitations of classical DT infrastructures by supporting quantum-resistant secure communication, intelligent task offloading, and low-latency processing in healthcare environments~\cite{jameil2025quantum, bera2025quantum}. In addition, quantum-inspired optimization and validation mechanisms, quantum cryptography integrated with blockchain-based DT frameworks, quantum machine learning techniques for privacy-preserving model training and diagnosis, and quantum networking solutions for reliable synchronization have been investigated to enhance the security, efficiency, and robustness of healthcare digital twin systems~\cite{selvagayathri2025next, prajapat2025blockchain, qu2023dtqfl, lv2022digital}. By integrating such quantum computing capabilities into DT pipelines, QDTs aim to enhance data processing, optimization, and decision-making for complex, data-intensive healthcare systems, although the practical realization of such integration introduces significant technical and system-level challenges.

In this paper, we review the key challenges that currently limit the practical deployment of QDTs in healthcare environments, with particular emphasis on constraints arising from quantum hardware limitations, hybrid classical–quantum integration, and cloud-based quantum computing. In addition, we outline research directions that we consider critical for enabling secure, reliable, and clinically trustworthy quantum digital twin systems. We argue that addressing these challenges is a necessary prerequisite before QDTs can be safely and effectively adopted in real-world healthcare applications.

The remainder of this paper is organized as follows. Section \ref{sec:DT} reviews digital twin technology in healthcare. Section \ref{sec:QDT} introduces QDTs and their underlying principles. Section \ref{sec:QDTforH} discusses quantum digital twin applications in healthcare, while Section \ref{sec:CHALLENGS} analyzes the key challenges that limit their practical deployment. Section \ref{sec:CALL} outlines future research directions and provides a call to action. Finally, Section \ref{sec:CONCLU} concludes the paper.

\section{Digital Twins for Healthcare}
\label{sec:DT}

\begin{figure}[t]
  \centering
  \includegraphics[width=\linewidth]{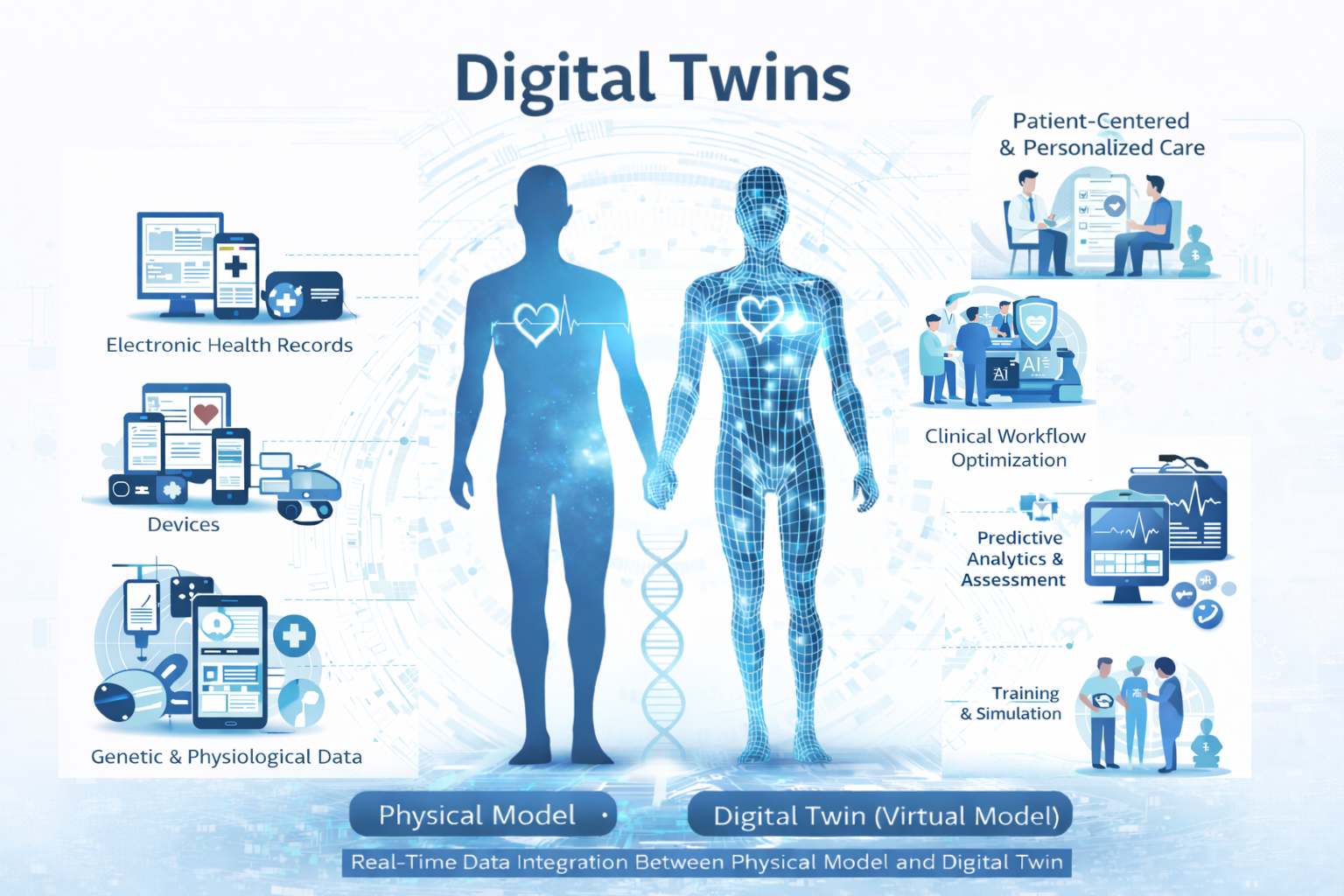}
  \caption{Conceptual overview of digital twin technology in healthcare, illustrating the interaction between the physical system (patient) and its digital twin through real-time data integration, and the main application domains including predictive analytics, clinical workflow optimization, and training and simulation.}
  \label{fig:dt-healthcare}
\end{figure}

Digital twin technology is transforming healthcare by integrating real-time data, advanced analytics, and virtual modeling to enable patient-centered and personalized care, support predictive analytics and preventive assessment, optimize clinical workflows, and facilitate training and simulation~\cite{vallee2023digital} (see Figure \ref{fig:dt-healthcare}). 
A summary of such usages is reported in Table~\ref{tab:tableSummary}, and they are discussed in the following.

By collecting patient data from sources such as electronic health records, medical devices, wearables, and genetic information, DTs provide a comprehensive view of patient conditions and support the development of personalized treatment plans based on individual characteristics and real-time physiological data~\cite{armeni2022digital, haleem2023exploring}. 
This personalized approach improves treatment effectiveness and patient satisfaction~\cite{goetz2018personalized}. DTs also support accurate and timely diagnosis by analyzing patient data, simulating diagnostic scenarios, and identifying patterns that may not be captured by traditional diagnostic methods, thereby reducing errors and enabling earlier intervention~\cite{venkatesh2022health, zhong2023overview}. Continuous monitoring through integration with wearable, remote monitoring, and IoT devices allows early detection of health deterioration, proactive intervention, and optimized care, particularly for patients with chronic conditions~\cite{elkefi2022digital, volkov2021digital}. In addition, DTs enhance patient engagement by providing access to personalized health information and treatment progress, supporting adherence to care plans and facilitating patient-centered decision-making~\cite{hassani2022impactful, abernethy2022promise, syed2023empowering}. The use of predictive analytics and machine learning enables the prediction of disease progression and treatment outcomes, supports early risk identification, and improves patient safety and long-term outcomes~\cite{allen2021digital}. Secure data sharing through DTs further ensures continuity of care and effective collaboration among healthcare teams~\cite{elkefi2022digital}.

Digital twin technology enables predictive analytics and preventive care by integrating real-time data with advanced analytical and machine learning techniques to identify health risks and anticipate disease progression~\cite{sun2023digital}. By analyzing comprehensive patient data, including medical history, lifestyle, genetic information, and physiological measurements, DTs can detect early risk indicators and support timely preventive interventions~\cite{kamel2021digital}. Through patient-specific modeling and analysis of historical trends, DTs can simulate disease evolution and forecast potential complications, allowing healthcare providers to adjust treatment strategies and implement targeted interventions to delay or prevent disease progression~\cite{sun2023digital, pascual2023systematic}. DTs support risk stratification by categorizing patients according to predicted risk levels, enabling efficient resource allocation and focused preventive care for high-risk individuals~\cite{coorey2022health, morande2022enhancing}. Continuous real-time monitoring further allows DTs to identify deviations from normal health parameters and trigger proactive clinical responses, reducing adverse events and improving patient outcomes. At a broader scale, the analysis of aggregated population data enables the identification of health trends and supports the design of targeted preventive strategies at the population level~\cite{popa2021use, calcaterra2023digital}.

By creating virtual representations of healthcare systems and integrating real-time data, DTs support the optimization of clinical operations through detailed analysis of workflows, resource utilization, and operational performance~\cite{venkatesh2022health}. Through the integration of data from electronic health records, medical devices, and administrative systems, DTs help identify bottlenecks, inefficiencies, and areas for process improvement, leading to more streamlined and efficient clinical workflows~\cite{armeni2022digital, haleem2023exploring}. Analysis of patient volumes, demand patterns, and resource utilization further supports effective allocation of staff, equipment, and facilities, while enabling capacity planning and informed decisions regarding future resource investments~\cite{elayan2021digital}. The use of predictive analytics and machine learning facilitates anticipation of operational scenarios such as patient flow and staffing requirements, allowing proactive adjustments to improve efficiency and patient care~\cite{van2022predictive}. DTs also support quality improvement and patient safety initiatives and facilitate communication and collaboration among healthcare teams through shared virtual platforms, contributing to improved care coordination and overall operational effectiveness~\cite{bruynseels2018digital, liu2019novel}.

Moreover, DTs provide important capabilities for training and simulation in healthcare by providing realistic virtual environments that replicate clinical systems and procedures, enabling skill development and decision-making without risk to patients~\cite{alazab2022digital, erol2020digital}. 
These environments can also be used during the development of medical devices. For example, by providing an accurate simulation of the patient's behavior, DTs may be used to test the software of a medical device in a risk-free manner (e.g.,~\cite{11058938}).
Additionally, these environments allow healthcare professionals, particularly surgeons, to rehearse procedures, evaluate different techniques, and improve technical proficiency through repeated practice~\cite{moztarzadeh2023metaverse}. A wide range of medical procedures, both invasive and non-invasive, can be simulated using DTs, allowing healthcare professionals to refine procedural skills and increase confidence prior to real clinical implementation~\cite{sun2023digital}. DTs are also valuable for emergency response training, as they can replicate critical scenarios such as trauma events and cardiac emergencies, supporting the development of rapid decision-making, teamwork, and coordination under high-pressure conditions~\cite{fan2021disaster}. By simulating complex clinical cases using patient data and real-time monitoring information, DTs enhance clinical reasoning and treatment planning in a safe environment~\cite{kaul2023role}. In addition, shared virtual platforms enable interdisciplinary collaboration and communication among healthcare professionals, improving coordination and integrated care delivery~\cite{iqbal2022use}. DTs further contribute to continuous professional development and research by supporting virtual training modules, case simulations, and experimental analysis of clinical and disease scenarios~\cite{semeraro2021digital}.

\begin{table*}[t]
\centering
\caption{Summary of application domains of DTs in healthcare.}
\label{tab:tableSummary}
\renewcommand{\arraystretch}{1.2}
\begin{tabular}{p{4cm} p{10.5cm}}
\hline
\textbf{Application Domain} & \textbf{Functions} \\
\hline

Patient-centered and personalized care 
& DTs enable individualized treatment planning, continuous monitoring, chronic disease management, patient engagement, and support for adherence and patient-centered decision-making based on integrated real-time health data~\cite{armeni2022digital, haleem2023exploring, goetz2018personalized, hassani2022impactful, syed2023empowering, elkefi2022digital}. \\

Predictive analytics and preventive assessment 
& DTs support diagnosis assistance, disease progression forecasting, treatment outcome prediction, risk stratification, early deterioration detection, preventive interventions, and population-level health trend analysis using advanced analytics and machine learning techniques~\cite{sun2023digital, pascual2023systematic, allen2021digital, venkatesh2022health, zhong2023overview, coorey2022health}. \\

Clinical workflow optimization 
& DTs enhance workflow analysis, bottleneck detection, resource allocation, capacity planning, staffing prediction, operational forecasting, quality improvement, and care coordination across healthcare systems~\cite{venkatesh2022health, armeni2022digital, elayan2021digital, van2022predictive, bruynseels2018digital}. \\

Training and simulation 
& DTs provides virtual environments for surgical rehearsal, procedure simulation, emergency response training, clinical reasoning development, interdisciplinary collaboration, medical device testing, and research experimentation without patient risk~\cite{alazab2022digital, erol2020digital, moztarzadeh2023metaverse, fan2021disaster, 11058938, kaul2023role}. \\

\hline
\end{tabular}
\end{table*}

\section{Quantum DTs}
\label{sec:QDT}
Digital twin technology originated in industrial engineering as a means of creating real-time virtual representations of physical systems for simulation, monitoring, analysis, and control. In recent years, DTs have gained increasing attention in domains such as healthcare, where continuous synchronization between physical entities and their digital counterparts enables predictive analytics, system optimization, and personalized decision-making. When quantum computing resources are integrated into digital twin architectures, \emph{Quantum Digital Twins} (QDTs) emerge, leveraging quantum algorithms and quantum processing units to enhance the analytical and operational capabilities of traditional digital twin models.

The role of quantum computing in DTs is both transformative and multifaceted. Quantum algorithms can significantly accelerate computational processes through quantum parallelism, enabling faster simulations and more efficient exploration of complex solution spaces. This capability is particularly valuable for applications involving engineering optimization, logistics planning, and supply chain management, as well as any aspects related with healthcare. Furthermore, quantum computing enhances data analysis by enabling more efficient handling of the massive datasets generated by DTs of complex systems, including urban environments and healthcare infrastructures~\cite{lin2022quantum}. 
QDTs also benefit from advances in quantum machine learning, which improve predictive analytics and adaptive modeling: Quantum-enhanced clustering, classification, and regression algorithms have demonstrated potential advantages over classical machine learning techniques for specific tasks, enabling faster anomaly detection, improved forecasting accuracy, and continuous model adaptation~\cite{liu2021rigorous, bartkiewicz2023synergic}.  From a security perspective, quantum technologies introduce advanced cryptographic mechanisms, including quantum-safe encryption schemes, which can significantly strengthen the protection of sensitive data exchanged between physical systems and their digital representations~\cite{xiao2022distributed,csafak2026quantum}.

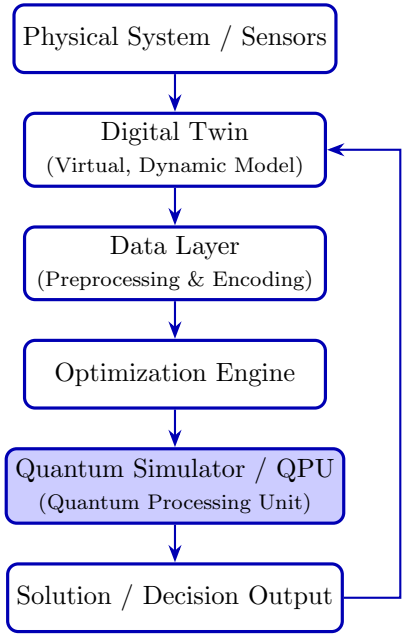
\begin{figure}[t]
  \centering
\usetikzlibrary{arrows.meta, positioning}
\begin{tikzpicture}[
    box/.style={
        rectangle,
        rounded corners,
        draw=blue!70!black,
        very thick,
        minimum width=4cm,
        minimum height=0.9cm,
        align=center,
        fill=white
    },
    arrow/.style={
        thick,
        ->,
        >=Stealth,
        draw=blue!70!black
    }
]

% Nodes
\node[box] (physical) {Physical System / Sensors};
\node[box, below=0.5cm of physical] (twin) {Digital Twin\\\footnotesize (Virtual, Dynamic Model)};
\node[box, below=0.5cm of twin] (data) {Data Layer\\\footnotesize (Preprocessing \& Encoding)};
\node[box, below=0.5cm of data] (opt) {Optimization Engine};
\node[box, fill=blue!20, below=0.5cm of opt] (quantum) {Quantum Simulator / QPU\\\footnotesize (Quantum Processing Unit)};
\node[box, below=0.5cm of quantum] (solution) {Solution / Decision Output};

% Forward arrows
\draw[arrow] (physical) -- (twin);
\draw[arrow] (twin) -- (data);
\draw[arrow] (data) -- (opt);
\draw[arrow] (opt) -- (quantum);
\draw[arrow] (quantum) -- (solution);

% Feedback loop
\draw[arrow] (solution.east) -- ++(0.75,0) |- (twin.east);

\end{tikzpicture}

  \caption{Conceptual architecture of a digital twin–quantum computing framework. Data flows from the physical system through the digital twin, data layer, optimization engine, and quantum simulator to generate a solution, with a closed-loop connection from the decision output back to the digital twin.}
  \label{fig:qdt-conceptual}
\end{figure}

A conceptual architecture of a quantum digital twin, inspired by~\cite{khumsikiew2025research}, is illustrated in Figure \ref{fig:qdt-conceptual}, where a classical digital twin is tightly coupled with a quantum computational layer through a data layer, an optimization engine, and a closed-loop feedback mechanism. %
In this architecture, the digital twin continuously collects real-time sensor data representing physical system states. These data are processed, normalized, and secured within the data layer before being forwarded to the quantum computational layer. The optimization engine translates digital twin models and system objectives into quantum-compatible problem formulations, which are then processed by a quantum simulator or quantum processor. The solutions generated by the quantum layer are fed back to the digital twin through a feedback loop, enabling real-time or near-real-time adaptation, recalibration, and decision-making~\cite{khumsikiew2025research}.

Several quantum algorithms play a central role in enhancing the computational capabilities of QDTs. Shor’s algorithm demonstrates the ability of quantum computing to efficiently factor large integers, highlighting its disruptive implications for cryptographic systems such as RSA and emphasizing the need for quantum-resistant security mechanisms in digital twin infrastructures~\cite{asthana2020decrypting, duan2022principles, xiao2022distributed}. As DTs become increasingly complex, the computational resources needed for accurate simulations also increase exponentially. The exponential speedup demonstrated by Shor’s algorithm highlights the potential of quantum computation for handling computationally intensive tasks.
Grover’s algorithm provides a quadratic speedup for searching unstructured databases, making it particularly relevant for data-intensive digital twin applications. By integrating Grover’s algorithm into digital twin analytics engines, systems can achieve faster real-time responses and improved predictive accuracy. This capability supports efficient analysis of large-scale datasets in data-intensive digital twin applications~\cite{seidel2023automatic}.
The quantum Fourier transform enables exponentially faster transformations between time and frequency domains compared to classical Fourier methods. In digital twin systems, this capability facilitates rapid analysis of large-scale sensor and signal data, supporting real-time simulation, optimization, and predictive maintenance across domains such as environmental monitoring, manufacturing, and healthcare~\cite{park2023reducing}.

Variational quantum algorithms, including the variational quantum eigensolver and the quantum approximate optimization algorithm, are particularly well suited for deployment on noisy intermediate-scale quantum devices. These algorithms are designed to tolerate hardware noise while solving complex optimization problems, making them practical candidates for near-term quantum digital twin implementations. Their integration into digital twin frameworks enables efficient system optimization, predictive maintenance, and operational planning in large-scale cyber–physical systems~\cite{lykov2023sampling, bermejo2023variational, bharti2022noisy}.

\section{Quantum DTs for Healthcare}
\label{sec:QDTforH}
Healthcare DTs provide patient-specific virtual models for simulating physiological processes, supporting improved diagnosis and personalized treatment planning~\cite{machado2023literature, abdollahi2022digital}, as well as platforms useful during the development of medical devices. Quantum computing introduces significant opportunities to further enhance healthcare DTs and personalized medicine. Quantum algorithms are capable of rapidly processing and analyzing large-scale datasets, thereby enabling DTs to capture a more comprehensive and individualized representation of patient health. For instance, quantum computing can efficiently analyze genomic data and integrate genetic information into patient-specific DTs, facilitating a more holistic approach to medical decision-making~\cite{ur2023quantum}. In addition to genetic data, quantum algorithms can evaluate a wide range of factors, including environmental conditions, lifestyle behaviors, and extensive medical histories, leading to more accurate and robust disease models~\cite{qu2023qnmf}. 
%
%\red{Andrea: Read and fixed till here}
%
These enhanced modeling capabilities are particularly impactful in oncology, where DTs can simulate tumor growth and progression based on patient-specific genetic mutations. Quantum computing can accelerate the simulation and assessment of chemotherapy and radiotherapy strategies, supporting the identification of treatment plans that achieve optimal therapeutic outcomes while minimizing adverse effects~\cite{kumela2023quantum}. The benefits of this approach extend to pharmacogenomics, where treatments can be customized according to an individual’s genetic profile, reducing the likelihood of adverse drug reactions and improving overall treatment effectiveness~\cite{upama2023quantum}.

\begin{figure}[t]
  \centering
  \includegraphics[width=0.5\linewidth]{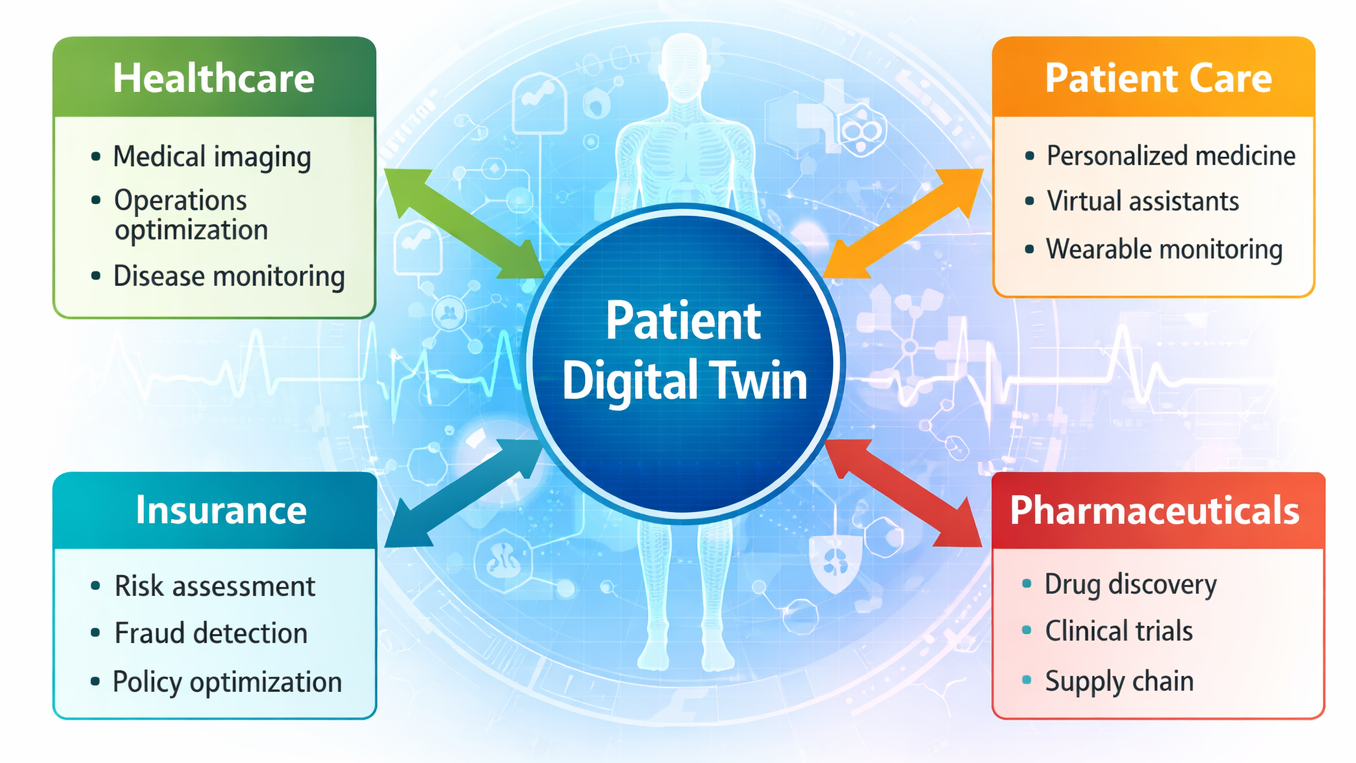}
  \caption{Conceptual illustration of the impact of QDTs across major healthcare sectors.}
  \label{fig:qdt-healthcare}
\end{figure}

More broadly, quantum computing has the potential to drive transformative advancements across multiple segments of the healthcare sector, including patient care, pharmaceutical development, hospital operations, and insurance services. Figure \ref{fig:qdt-healthcare} provides an overview of the transformative impact of quantum computing across key segments of the healthcare ecosystem~\cite{khumsikiew2025research}.
In patient care, quantum-enhanced analytics enable the efficient processing of large genomic datasets, supporting highly personalized treatment plans and advancing the practice of precision medicine. Furthermore, quantum-powered artificial intelligence–based virtual health assistants could deliver real-time, individualized medical guidance, significantly improving patient experience and clinical outcomes. Quantum algorithms may also enhance wearable health technologies by enabling more accurate analysis of real-time physiological data and improving the prediction of critical health events, such as cardiac incidents.
In the pharmaceutical domain, quantum computing is expected to revolutionize drug discovery and clinical trial processes. By simulating molecular interactions at unprecedented scales, quantum algorithms can facilitate the identification and optimization of drug candidates more efficiently than classical approaches. Additionally, quantum-enhanced data analysis can support faster and more accurate evaluation of clinical trial results, thereby reducing the time required to bring new therapies to market.

Hospitals are also poised to benefit from quantum computing advancements. Quantum algorithms can improve diagnostic accuracy and efficiency in medical imaging techniques, including magnetic resonance imaging and computed tomography. Beyond diagnostics, quantum computing can optimize hospital operations, such as staff scheduling and resource allocation, leading to improved efficiency and patient care quality. Moreover, quantum-enhanced data analysis can assist in monitoring patient information to detect and mitigate potential disease outbreaks within healthcare facilities, contributing to proactive and preventive healthcare management.
Insurance systems represent another area where quantum computing can have a substantial impact. Quantum algorithms can analyze complex, multi-dimensional datasets to provide more refined assessments of individual risk profiles. This capability supports more accurate pricing of insurance policies and enhances the detection of fraudulent healthcare claims. Furthermore, quantum computing can assist in designing insurance plans that balance cost-effectiveness for providers with comprehensive coverage for policyholders by analyzing population health trends and historical policy data.

It should be noted that the practical realization of healthcare-oriented QDTs critically depends on secure data exchange and low-latency computation. Healthcare DTs rely on continuous transmission of sensitive physiological data across IoT devices, edge nodes, and cloud platforms, making them vulnerable to advanced cyber threats and communication delays. Recent studies~\cite{jameil2025quantum, bera2025quantum} on quantum-resistant secure communication and quantum-assisted task offloading demonstrate that classical DT infrastructures are insufficient to meet the stringent security and real-time requirements of healthcare environments, thereby motivating the integration of quantum-aware security mechanisms and intelligent workload distribution strategies in future healthcare digital twin systems. Quantum-inspired approaches have been applied to digital twin frameworks to accelerate optimization and validation processes, enabling faster encryption, reduced validation latency, and improved resilience against emerging quantum threats in real-time healthcare DT environments~\cite{selvagayathri2025next}. Blockchain-enabled digital twin networks have further incorporated quantum cryptography to ensure data confidentiality and integrity, addressing the limitations of classical encryption when synchronizing sensitive patient data between physical and virtual twins~\cite{prajapat2025blockchain}. Quantum computing has also been integrated with DTs through variational quantum neural networks and federated learning to support privacy-preserving, fast model training and real-time intelligent diagnosis~\cite{qu2023dtqfl}. Also, quantum networking techniques based on quantum key distribution and entanglement-assisted communication have been explored to enable secure and reliable synchronization between physical assets and their DTs under high network load~\cite{lv2022digital}.

\section{Key Challenges in Quantum DTs for Healthcare}
\label{sec:CHALLENGS}
Despite their potential benefits, the practical deployment of QDTs for healthcare remains highly challenging. These challenges primarily arise from the limitations of current quantum hardware, the complexity of integrating hybrid classical–quantum architectures, and the stringent requirements of healthcare environments, where reliability, low latency, security, and regulatory compliance are essential. Consequently, a systematic understanding of the key challenges associated with QDTs is necessary before such systems can be adopted in real-world healthcare settings.
%-Quantum Hardware Limitations (NISQ Era)
%-Hybrid Classical–Quantum Integration Complexity
%-Latency and Cloud-Based Quantum Access
%-Security and Trust Challenges

Current QDT implementations are constrained by the limitations of noisy intermediate-scale quantum (NISQ) hardware, including limited qubit counts, short coherence times, and high gate error rates. These hardware constraints restrict the size and depth of executable quantum circuits and limit the range of healthcare-related optimization and simulation tasks that can be reliably accelerated using quantum resources, thereby reducing the practical impact of QDTs in safety-critical medical applications~\cite{preskill2018quantum, hassija2020present}.

A fundamental challenge in the development of QDTs lies in the efficient sampling of high-dimensional probability distributions, which is essential for modeling complex real-world systems~\cite{amir430446can}. While quantum computation naturally produces samples from probability distributions through measurement, identifying suitable unitary operators that can be implemented as polynomial-depth quantum circuits remains an open problem~\cite{amir430446can}. Moreover, encoding complex system state spaces into binary representations suitable for quantum processing often requires a large number of qubits, which is incompatible with the scale of current quantum hardware. As a result, despite the theoretical relevance of quantum computing for probabilistic modeling, unresolved algorithmic and scalability constraints significantly limit the practical applicability of QDTs~\cite{amir430446can}.

Moreover, QDTs inherently depend on hybrid classical–quantum architectures, which introduce significant integration complexity. Classical DT components are responsible for data acquisition, preprocessing, and control, while quantum processors are selectively used as computational accelerators. The repeated encoding of classical healthcare data into quantum representations and the decoding of quantum outputs introduce additional overhead, synchronization challenges, and potential loss of information fidelity, particularly in dynamic healthcare environments that require frequent model updates and near-real-time responses~\cite{das2019case, niu2023enabling}.

Another critical challenge arises from the reliance on cloud-based quantum computing infrastructures. Due to the high cost and operational complexity of quantum hardware, access to quantum processing units is predominantly provided through remote and shared cloud platforms. This execution model introduces communication latency, queueing delays, and limited control over execution timing, which can conflict with the strict low-latency requirements of healthcare DTs, such as continuous patient monitoring and emergency decision support systems~\cite{das2019case, kilber2021cybersecurity}.

Security and trust issues further complicate the deployment of QDTs in healthcare settings. Studies on shared quantum computing environments have demonstrated that multi-tenant execution can expose quantum workloads to new classes of vulnerabilities, including crosstalk-induced interference, adversarial circuit manipulation, and information leakage across concurrently executed programs. Such vulnerabilities undermine the integrity and reliability of quantum-assisted computations and raise serious concerns for healthcare applications, where erroneous or manipulated outputs may directly impact clinical decisions and patient safety~\cite{ovaskainen2025quantum, sarker2026machine}.

In addition to security concerns, scalability remains a fundamental challenge for healthcare-oriented QDTs. Quantum computing resources are scarce, expensive, and typically shared among multiple users through multi-tenant cloud platforms~\cite{ovaskainen2025quantum}. As healthcare DTs scale to support multiple patients, devices, and clinical workflows, efficient resource allocation, workload scheduling, and isolation become increasingly difficult. Current quantum infrastructures lack mature mechanisms to prioritize healthcare-critical tasks while ensuring fairness and performance stability across concurrent workloads, further limiting the feasibility of large-scale quantum digital twin deployments.

Beyond technical and infrastructural challenges, establishing clinical trust in QDT outputs represents a major barrier to adoption. Healthcare decision-making requires results that are not only accurate but also interpretable, reproducible, and clinically verifiable. The probabilistic nature of quantum computation, combined with hardware noise and variability across executions, complicates result validation and explainability. Without robust mechanisms for clinical validation and transparent interpretation of quantum-assisted outputs, healthcare professionals may be reluctant to rely on QDTs for safety-critical decisions.

\section{Call to Action}
\label{sec:CALL}
To address the challenges hindering the adoption of QDTs in healthcare environments, we urge the research community and healthcare system designers to focus on the following research directions:

\begin{itemize}

 \item Deploy a secure quantum software stack~\cite{dwivedi2024quantum,ovaskainen2025quantum} for QDTs:
To realize the full potential of QDTs, the research community must prioritize the deployment of a secure quantum software stack for QDTs. Each layer of the stack must address both the high-fidelity synchronization of the QDT and the cryptographic security required for quantum-classical communication.

 \item Healthcare-aware quantum-classical architectures for QDTs:
Designing hybrid quantum-classical digital twin architectures that explicitly account for healthcare constraints such as low latency, real-time monitoring, and safety-critical decision-making remains a priority. Researchers should focus on clearly defining which parts of digital twin sub-tasks are suitable for quantum acceleration under current NISQ era technology and within shared cloud-access environments.

 \item Quantum algorithmic feasibility:
There is a need for systematic investigation into quantum algorithms that have potential to be realistically implemented for QDTs workloads.

%%%%%%%%%%%%%%%%%%%%%
 \item Security and isolation in multi-tenant quantum environments:
Strengthening security for QDTs requires improved isolation mechanisms, protection against multi-tenant interference, crosstalk~\cite{sarker2026machine}, and robust handling of shared quantum resources, particularly when sensitive healthcare data and clinical decision processes are involved.
%%%%%%%%%%%%%%%%%%%%
 \item Validation, reproducibility, and clinical trust for QDTs:
Establishing standardized validation and benchmarking frameworks for quantum-assisted digital twin outputs is a must. Collaboration between quantum researchers and healthcare professionals is required to align technical validation with medical standards.

\end{itemize}

\section{Conclusion}
\label{sec:CONCLU}

This paper examined the integration of quantum computing into healthcare digital twin systems and highlighted the key challenges that currently hinder the practical deployment of Quantum Digital Twins. While digital twins already play an important role in healthcare applications such as monitoring, decision support, and system optimization, their growing scale, data intensity, and real-time requirements increasingly stress classical computing infrastructures.
Quantum computing offers promising capabilities to enhance selected digital twin functions, particularly for computation-intensive and security-sensitive tasks. However, this work shows that the integration of quantum resources into healthcare digital twin pipelines is far from straightforward. Limitations of current quantum hardware, the complexity of hybrid classical–quantum system design, cloud-based quantum access, and the need to meet strict healthcare constraints collectively represent major barriers to adoption. The findings of this study indicate that the successful evolution of Quantum Digital Twins will require careful, challenge-aware system design. Future progress depends on incremental advances in quantum hardware, improved strategies for hybrid classical–quantum integration, and the establishment of validation and trust mechanisms aligned with clinical practice. Addressing these challenges, as outlined in the call to action, is a necessary step toward developing quantum digital twin systems that are technically feasible, reliable, and clinically trustworthy for real-world healthcare use.

\section*{Acknowledgement}
This work has been supported by the National Recovery and Resilience Plan (NRRP), Mission 4, Component 2, Investment 1.1, under Call No. 104 (2.2.2022) by the Italian Ministry of University and Research (MUR), funded by the EU – NextGenerationEU – SAFEST project (CUP F53D23004230006), Grant Decree n. 861 (16-6-2023) by MUR.
This work was supported in part by project SERICS (PE00000014) under the NRRP MUR program funded by the EU - NGEU. 
The work of Andrea Bombarda has been supported by PNRR - ANTHEM (AdvaNced Technologies for Human-centrEd Medicine) - Grant PNC0000003 – CUP: B53C22006700001 - Spoke 1 - Pilot 1.4.
M.H. acknowledges support from the Business Finland through project \textit{SeQuSoS} (Grants No. 112/31/2024), and Research Council of Finland through project \textit{Profi 8} (Grants No. 365343).

\bibliographystyle{plain}
\bibliography{main}

\end{document}